\documentclass[aps,prl,10pt,twocolumn,showpacs,preprintnumbers,amsmath,amssymb,superscriptaddress]{revtex4-1}
\usepackage[english]{babel}
\usepackage[dvips]{graphicx}
\usepackage{dcolumn}
\usepackage{bm}

\usepackage{color}

\begin{document}

\title{Expansion of a radially symmetric blast shell into a uniformly magnetized plasma}

\author{M. E. Dieckmann}
\affiliation{Department of Science and Technology (ITN), Link\"opings University, Campus Norrk\"oping, SE-60174 Norrk\"oping, Sweden}
\author{Q. Moreno}
\affiliation{Universit\'e de Bordeaux, CNRS, CEA, CELIA, Talence, France}
\author{D. Doria}
\affiliation{Centre for Plasma Physics (CPP), Queen's University Belfast, BT7 1NN, UK}
\author{L. Romagnani}
\affiliation{\'Ecole Polytechnique, CNRS, LULI, F-91128 Palaiseau, France}
\author{G. Sarri}
\affiliation{Centre for Plasma Physics (CPP), Queen's University Belfast, BT7 1NN, UK}
\author{D. Folini}
\affiliation{\'Ecole Normale Sup\'erieure, Lyon, CRAL, UMR CNRS 5574, Universit\'e de Lyon, France}
\author{R. Walder}
\affiliation{\'Ecole Normale Sup\'erieure, Lyon, CRAL, UMR CNRS 5574, Universit\'e de Lyon, France}
\author{A. Bret}
\affiliation{ETSI Industriales, Universidad de Castilla-La Mancha, 13071 Ciudad Real, Spain}
\affiliation{Instituto de Investigaciones Energeticas y Aplicaciones Industriales, Campus
Universitario de Ciudad Real, 13071 Ciudad Real, Spain}
\author{E. d'Humi\`eres}
\affiliation{Universit\'e  de Bordeaux, CNRS, CEA, CELIA, Talence, France}
\author{M. Borghesi}
\affiliation{Centre for Plasma Physics (CPP), Queen's University Belfast, BT7 1NN, UK}

\email{mark.e.dieckmann@liu.se}

\begin{abstract}
The expansion of a thermal pressure-driven radial blast shell into a dilute ambient plasma is examined with two-dimensional PIC simulations. The purpose is to determine if laminar shocks form in a collisionless plasma that resemble their magnetohydrodynamic counterparts. The ambient plasma is composed of electrons with the temperature 2 keV and cool fully ionized nitrogen ions. It is permeated by a spatially uniform magnetic field. A forward shock forms between the shocked ambient medium and the pristine ambient medium, which changes from an ion acoustic one through a slow magnetosonic one to a fast magnetosonic shock with increasing shock propagation angles relative to the magnetic field. The slow magnetosonic shock that propagates obliquely to the magnetic field changes into a tangential discontinuity for a perpendicular propagation direction, which is in line with the magnetohydrodynamic model. The expulsion of the magnetic field by the expanding blast shell triggers an electron-cyclotron drift instability. 
\end{abstract}

\maketitle

\section{Introduction}

The expansion of a blast shell into an ambient plasma triggers the formation of shocks if the blast shell front moves faster than the relevant charge density wave in the ambient plasma. The properties of the shock depend on many factors; among others on how important binary collisions between the particles are for the plasma evolution, on the ratio $\beta$ between the plasma's thermal to magnetic pressure, on the magnetic field direction, on how the electron temperature compares to that of the ions and on the charge state and composition of the ions. 

The probably simplest form of a magnetized shock in plasma is that obtained from a single-fluid MHD model. It is an appropriate description of magnetized shocks in plasma, in which the collision frequency between particles exceeds by far the characteristic frequencies of all processes that are relevant for the shock's evolution. Shocks in this approximation can be subdivided into magnetized shocks, which are mediated by the fast or slow magnetosonic waves, and into hydrodynamic shocks that propagate approximately parallel to the magnetic field. 

It is interesting to determine the degree to which these shocks exist in other plasma models, such as the kinetic model, and how the selected approximation affects the shock's properties. Magnetized collision-less shocks can be driven in particle-in-cell (PIC) simulations by a magnetic pressure gradient \cite{Forslund71} or by a drifting plasma \cite{Lembege01}. Another way is to let a rarefaction wave \cite{Sack87,Grismayer06,Grismayer08,Thaury10,Dieckmann12}, which is driven by a sharp change in the thermal pressure, collide with an ambient plasma. 

The latter case was studied in Ref. \cite{Dieckmann16} with one-dimensional PIC simulations and in the presence of an initially spatially uniform perpendicular magnetic field. The jump in the thermal pressure between both plasmas accelerated the dense plasma. A rarefaction wave propagated into the dense plasma and accelerated the ions into the opposite direction. The mean speed of the accelerated ions increased linearly with the distance from the front of the rarefaction wave. The ions' peak speed was limited by the thermal and magnetic pressure of the ambient medium that resisted the blast shell's expansion. A shock developed between the blast shell's front and the ambient plasma, which was initially mediated by quasi-electrostatic lower-hybrid waves. The ambient plasma, which crossed the shock, piled up behind it and formed a hot plasma population; the shocked ambient medium. The shock propagated at a speed that exceeded the lower-hybrid speed but remained well below that of the fast magnetosonic wave at lower frequencies. The ambient magnetic field was depleted in the rarefaction wave and it piled up close to the shock. 

Reference \cite{Dieckmann17} followed the plasma expansion over a longer time. The lower-hybrid shock changed into a fast magnetosonic shock on a time scale of the order of tens of inverse lower-hybrid frequencies. The frequencies of the shocked fast magnetosonic waves were slightly below the lower-hybrid frequency, where the fast magnetosonic waves coupled to the lower-hybrid wave branch. The phase speed of the waves decreased with increasing wavenumbers in this frequency band and the shock became dispersive. Consequently the shock changed into a train of lower-hybrid solitons. A tangential discontinuity grew that separated the front of the blast shell from the shocked ambient medium. The shock speed in Ref. \cite{Dieckmann17} equalled 1.5 times the fast magnetosonic speed. Shocks with such a low Mach number reflect only a small fraction of the inflowing upstream ions. The beam they formed was not energetic enough to enforce the cyclic shock reformation, which is observed for collisionless magnetized shocks with a higher Mach number \cite{Lembege92,Scholer92,Shimada00,Hoshino02,Scholer03,Lee05,Chapman05,Burgess07,Marcowith16,Gueroult17,Schaeffer17}. 

The subdivision of the plasma into a rarefaction wave, into a tangential discontinuity that separated the blast shell plasma from the shocked ambient plasma and a laminar forward shock observed in Ref. \cite{Dieckmann17} closely followed the plasma distribution we would expect from a MHD model. The shock formed on time scales much shorter than an inverse ion gyro-frequency because it involved the high-frequency part of the fast magnetosonic mode. It is of significant interest to explore how this shock changes as the angle between the shock normal and the magnetic field direction is altered; not only from a theoretical point of view but also with respect to forthcoming experiments similar to that in Ref. \cite{Schaeffer17}. Such experiments, in which a blast shell of collisionless plasma is created by the ablation of a solid target by an intense laser pulse and interacts with a second plasma population, allow us to study in the laboratory processes that take place in energetic astrophysical or solar system plasma \cite{Remington99,Zakharov03}. A related experiment has studied the release of ion clouds by the AMPTE satellite mission \cite{Bernhardt87}, which led to the formation of shock-like structures \cite{Chapman89}.

We study here with PIC simulations, which resolve 2 spatial and 3 velocity dimensions, the expansion of an initially radially symmetric blast shell of collisionless plasma into a magnetized ambient medium. The magnetic field in one simulation is aligned with one of the directions resolved by the simulation plane, which allows us to study the formation of shocks for a wide range of angles between the shock normal and the magnetic field direction. A second simulation considers a magnetic field, which is aligned with the normal of the simulation plane. We test with this simulation if instabilities like the Weibel instability in an unmagnetized rarefaction wave \cite{Weibel59, Quinn12} develop also in a magnetized rarefaction wave. Our paper is structured as follows. Section 2 summarizes the equations, which are solved by the PIC code, and the initial conditions. Section 3 presents the results, which are summarized in section 4. 

\section{Algorithm and initial conditions}

A PIC code represents the electric field $\mathbf{E}$ and the magnetic field $\mathbf{B}$ on a numerical grid. Both fields are evolved in time using Amp\`ere's law and Faraday's law.
\begin{equation}
\mu_0 \epsilon_0 \frac{\partial \mathbf{E}}{\partial t} = \nabla \times \mathbf{B} - \mu_0 \mathbf{J}, \label{Ampere}
\end{equation}
\begin{equation}
\frac{\partial \mathbf{B}}{\partial t} = - \nabla \times \mathbf{E}. \label{Faraday}
\end{equation}
The vacuum permittivity and permeability are $\epsilon_0$ and $\mu_0$. Gauss' law and $\nabla \cdot \mathbf{B}=0$ are fulfilled to round-off precision by the EPOCH code \cite{Arber15,Esirkepov01}. Each plasma species $i$ is represented by one phase space density distribution $f_i(\mathbf{x},\mathbf{v},t)$, which is approximated by an ensemble of computational particles (CPs) with a charge-to-mass ratio $q_i/m_i$ that equals that of the plasma species it represents. Their velocities are updated with the Lorentz force equation and the electromagnetic fields, which have been interpolated from the grid to the particle position. The position is updated with its velocity and the simulation time step $\Delta_t$. We interpolate the current density of each CP of species $i$ to the grid and sum up the contributions of all CPs, which gives $\mathbf{J}_i$. The global current $\mathbf{J} = \sum_i \mathbf{J}_i$ is used to update the electric field with Amp\`ere's law. 

\begin{figure}
\includegraphics[width=0.8\columnwidth]{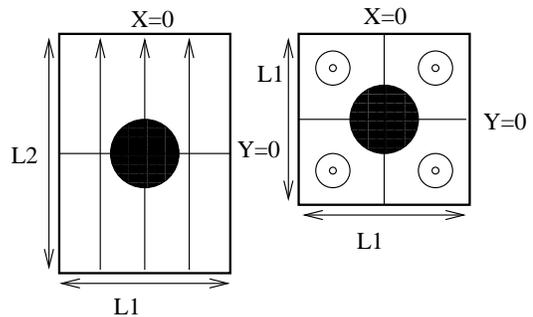}
\caption{The initial plasma distributions. The black circle shows the location of the dense ions with the density 15$n_0$ in the interval with radius $r \le r_0$, which is surrounded by the ambient ions with the density $n_0$. Panel (a) sketches the box of simulation 1 with the side length $L_1=4r_0$ along x and with the side length $L_2=2L_1$ along y. The magnetic field is aligned with y. Panel (b) shows the simulation box geometry of simulation 2. The size of the simulation box is $L_1$ along x and y and the magnetic field is aligned with z. The boundary conditions are periodic in all directions.}\label{Sketch}
\end{figure}

We model fully ionized nitrogen ions and electrons with the correct mass ratio $m_n/m_e \approx 2.6 \times 10^4$ and with the correct electron charge-to-mass ratio $e/m_e$. The plasma is distributed in the two-dimensional simulation plane according to Fig. \ref{Sketch}.  
Simulation 1 resolves the interval $L_1 = 16$ mm by 2000 grid cells and $L_2=32$ mm by 4000 grid cells. Simulation 2 resolves the side length of the quadratic box by 2000 grid cells. The origin of the coordinate system is placed in the center of the simulation box and the radius is $r = \sqrt{x^2+y^2}$. A circular boundary with radius  $r_0=2$ mm separates a dense plasma in the interval $r \le r_0$ from the ambient plasma with $r>r_0$. The azimuth angle relative to the positive y-axis in the counter-clockwise direction is $\rho$. 

The ambient plasma consists of ions with the number density $n_0 = 1.42 \times 10^{14}\textrm{cm}^{-3}$. The ions are 15 times denser in the interval $r\le r_0$. The electrons are 7 times denser than the ions. The electron temperature of the ambient medium is $T_0 = 2$ keV giving an electron thermal speed $v_{te}={(k_BT_0/m_e)}^{1/2} \approx 1.9 \times 10^7$ m/s ($k_B:$ Boltzmann constant). The electron plasma frequency $\omega_{pe}={(7e^2n_0/m_e \epsilon_0)}^{1/2}$ is $\approx 1.8 \times 10^{12}\, \mathrm{s}^{-1}$ in the ambient plasma and the ion plasma frequency $\omega_{pi}={(7m_e/m_n)}^{1/2}\omega_{pe} \approx 3 \times 10^{10}\,\mathrm{s}^{-1}$.

On average, thermal diffusion lets electrons stream from the dense into the dilute plasma. Consequently,  the dense plasma will go onto a positive potential relative to the dilute one. Electrons, which enter the dense plasma, are accelerated by this potential jump and form a beam of energetic electrons. We set the electron temperature within the dense cloud to $2T_0$, which suppresses two-stream instabilities between this beam and the thermal electrons. The ion temperature is $T_0/12.5$ everywhere. 

Simulation 1 resolves the electrons by $6.4\times 10^8$ computational particles (CPs) and the ions by $9.6 \times 10^8$ CP's. Simulation 2 employs one half of the total number of CPs. One half of the CPs is placed in the interval $r\le r_0$.  Initially the net charge and current vanish everywhere in the simulation box. The electric field and all magnetic field components except that of the background magnetic field are set to zero at the time $t=0$. The background magnetic field has the amplitude $B_0=$ 0.85 T in both simulations, which gives a value of $\beta \equiv (7n_0k_BT_0)/(B_0^2/2\mu_0)= 1.1$ in the ambient plasma, where we neglected the pressure contribution of the cool ions. It is aligned with y in simulation 1 and with z in simulation 2. The ambient electron's thermal gyro-radius $r_{ge}=v_{te}/\omega_{ce}$ is $r_{ge}=0.125$ mm ($\omega_{ce}=eB_0/m_e$). 

The ion acoustic speed in the ambient plasma is $c_s \approx 4 \times 10^5$ m/s with $c_s={((\gamma_ek_BT_0+\gamma_nk_BT)/m_n)}^{1/2}$. We assumed that the adiabatic constants of electrons and ion are $\gamma_e = 5/3$ and $\gamma_n=3$, respectively. The Alfv\'en speed $v_A = B_0/{(\mu_0 n_0 m_n)}^{1/2}$ is $v_A\approx 4.1 \times 10^5$ m/s for our plasma parameters. The speed of the fast magnetosonic wave $v_{fms}={(c_s^2+v_A^2)}^{1/2}$ for perpendicular propagation is $v_{fms}\approx 5.8\times 10^5$ m/s. 

We can estimate the speeds of density waves for intermediate angles using a one-fluid MHD model that is valid at frequencies below the ion gyrofrequency, which is in our case that of fully ionized nitrogen $\omega_{ci}=7eB_0/m_n$. The sound speed 
$\tilde{c}_s$ in the collisional MHD plasma is close to the ion acoustic speed $c_s$ in collisionless plasma and the same holds for the fast magnetosonic speed. Linear Alfv\'en waves, which propagate along the magnetic field, can not compress the plasma and $\tilde{c}_s$ is the only relevant phase speed of density waves that propagate along this direction. The fast magnetosonic speed is the only relevant one in the MHD plasma if the density waves propagate perpendicularly to the magnetic field. Sound waves, which propagate obliquely to the magnetic field, can change into slow magnetosonic modes. Slow- and fast magnetosonic modes coexist for a wide range of oblique propagation angles and the phase speed $v_f$ ($v_s$) of the fast (slow) obliquely propagating magnetosonic mode is 
$$
\frac{2v^2_{f,s}}{v^2_A} = (1+\tilde{\beta}) \pm {\left ( {(1-\tilde{\beta})}^2 + 4 \tilde{\beta} {(\sin \rho)}^2 \right )}^{1/2},
$$ 
where $\tilde{\beta} = \tilde{c}_s^2 / v_A^2$. An in-depth discussion of magnetosonic modes in MHD plasma and in kinetic plasma and related observations can be found in \cite{Verscharen17}. We can identify these modes in the PIC simulation with the phase relation between the plasma density and the magnetic amplitude. Oscillations of the plasma density and magnetic field are in phase in the case of the fast magnetosonic wave and in antiphase for slow magnetosonic modes. 

\section{The simulation results}

We consider the plasma distribution at the time $t_{sim}=$ 3.57 ns or $t_{sim} \omega_{ci}/2\pi \approx 2.3 \times 10^{-2}$. If we observe shocks, then they are mediated by gradients in the thermal and magnetic pressures and not by a Larmor rotation of the upstream ions in the downstream magnetic field. All densities are normalized to the ion density $n_0$ of the ambient medium and the magnetic pressure is normalized as $P_B(x,y)=(B_x^2(x,y)+B_y^2(x,y)+B_z^2(x,y))/B_0^2$.

\subsection{Simulation 1: magnetic field aligned with y}

\begin{figure*}
\includegraphics[width=\textwidth]{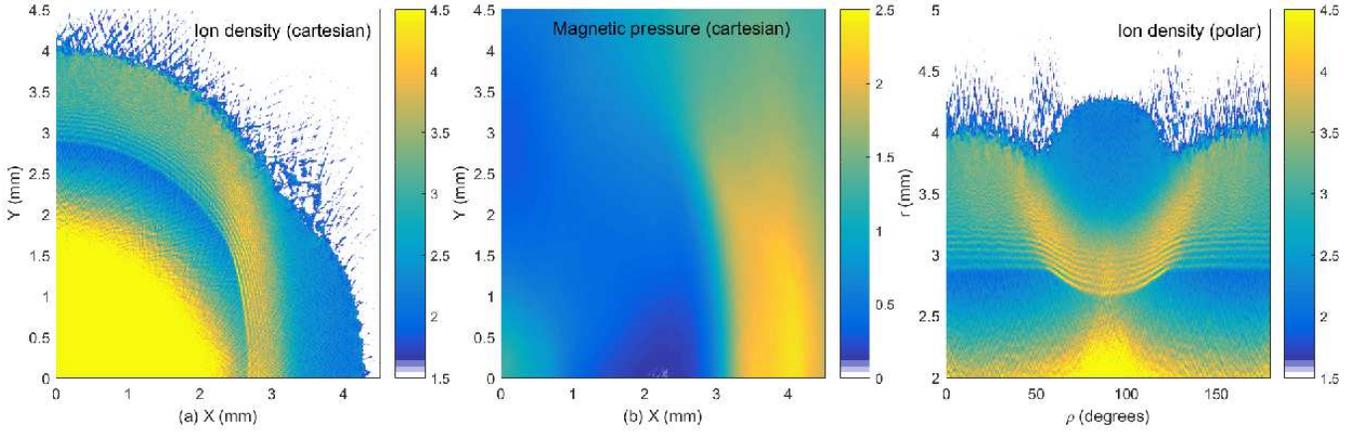}
\caption{The distribution of the ion density $n_i$ and of the normalized magnetic pressure $P_B=(B_x^2+B_y^2+B_z^2)/B_0^2$. The ion density is shown in cartesian coordinates in (a). Panel (b) shows the magnetic pressure in cartesian coordinates. The ion density distribution in polar coordinates is shown in panel (c). The linear color scale is clamped to the value 1.5 in (a, c) in order to emphasize the location of the shock. The time is $t_{sim}=3.57$ ns (Multimedia view).}\label{figure2}
\end{figure*}

Figure \ref{figure2}(a) shows the ion density in the quadrant $x>0$ and $y>0$. The ion density along the axis $x=0$ decreases below 4.5, which is the maximum value displayed on the color scale, at $y\approx 2$ mm, it reaches its minimum value $\approx 2$ at $y\approx$ 2.8 and increases to over 3 just behind the shock, which is located at $y\approx 4$ mm. This density profile resembles that of a circular blast shell in unmagnetized plasma \cite{Dieckmann17b}. The ion density distribution maintains a radially symmetric profile up to $x\approx $ 2 mm. A striped high density band is located in the interval 2.7 mm $\le x \le 3.2$ mm and $y \le$ 2 mm. 

Figure \ref{figure2}(b) shows that the front of this high-density band coincides with an interval with a steep gradient of $P_B$ up to $y\approx 3$ mm. The front of the perpendicular shock in Fig. \ref{figure2}(a) is located at $x\approx 4.3$ mm for $y\approx 0$ and the magnetic pressure in the interval between the shock and the high-density band is higher than that upstream. A shock, which compresses the plasma density and the magnetic pressure like the one moving along $\rho \approx 90^\circ$, is mediated by the fast magnetosonic mode. 

Figure \ref{figure2}(c) shows the ion density distribution in polar coordinates. The shock front is fastest and its separation from the trailing high-density band is largest for $\rho \approx 90^\circ$. The radius of the latter increases as we move away from $\rho = 90^\circ$ until $\rho \approx 60^\circ$ or $\rho \approx 120^\circ$, which is what we expect from Fig. \ref{figure2}(a) since there the high-density structure is field-aligned. A second high-density structure extends in the direction $\rho = 90^\circ$ up to $r\approx 2.5$ mm. The gradients of the magnetic pressure and of the thermal pressure are parallel along this direction, as we can see from Fig. \ref{figure2}, which causes a stronger acceleration of the blast shell ions in this direction. Figure \ref{figure2}(c) reveals that the density stripes, which were also seen in the ion high-density band in Fig. \ref{figure2}(a), are continuous for angles that range from $\rho = 0$ to $\rho = 90^\circ$. 

\begin{figure}
\includegraphics[width=\columnwidth]{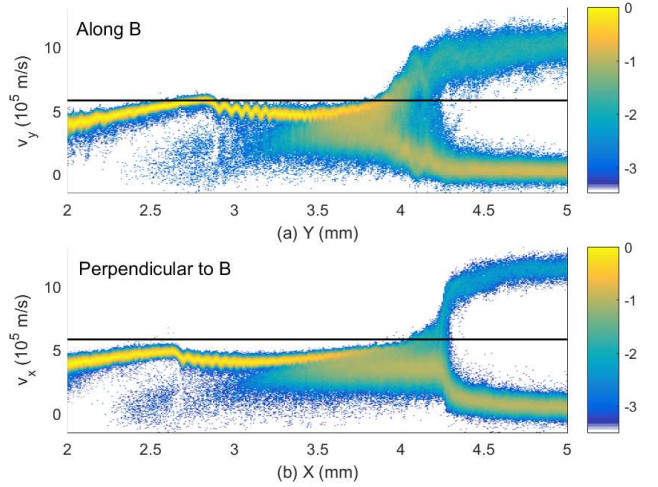}
\caption{Ion phase space density distribution at $t_{sim}=3.57$ ns: panel (a) shows the distribution $f_n (y,v_y)$ along y, which has been averaged over the interval -0.2 mm $\le x \le $ 0.2 mm. It depicts an ion acoustic shock that propagates along the background magnetic field. Panel (b) shows the distribution $f_n (x,v_x)$ along x, which has been averaged over the interval -0.2 mm $\le y \le $ 0.2 mm. It is a fast magnetosonic shock that propagates perpendicularly to the baclground magnetic field. The color scale is 10-logarithmic. Horizontal line: fast magnetosonic speed $5.8 \times 10^5$ m/s (Multimedia view).}\label{figure3}
\end{figure}

According to Fig. \ref{figure2}(c) these stripes are located in the intervals 2.9 mm $\le r \le $ 3.2 mm for $\rho = 0$ and 2.6 mm $\le r \le $ 3.1 mm for $\rho = 90^\circ$. Figure \ref{figure3} reveals their cause by looking at the shocks that flow along and perpendicular to the magnetic field. The density stripes correspond to ion density waves, which cause velocity oscillations of the blast shell ions. They are ion acoustic waves for $\rho = 0$ and lower-hybrid waves for $\rho=90^\circ$. The oscillations start at the locations, where the blast shell ions are no longer accelerated by the electric field of the rarefaction wave. These positions are $y=2.9$ mm and $v_y \approx 6 \times 10^5$ m/s for the unmagnetized shock and $x=2.7$ mm and $v_x \approx 4.5 \times 10^5$ m/s for the magnetized shock. The oscillation amplitudes of the mean velocity and of the density decrease with an increasing positive distance from these positions. These oscillations resemble those found at the boundary between a hydrodynamic rarefaction wave and the velocity plateau \cite{Gurevich84}. Their cause is the discontinuous first derivative of the mean velocity. 

The mean speed of the blast shell ions is lower for $\rho = 90^\circ$ than for $\rho = 0$, which explains why the density stripes in Fig. \ref{figure2}(c) are located at lower radii. The amplitude of the density modulations changes with $\rho$ because they are tied to different wave modes.
Comparing the location of the shocks and the distributions of the downstream ions in Fig. \ref{figure3} shows that, even though the blast shell ions and the ions of the ambient medium behind the shock with $\rho = 90^\circ$ are propagating at a lower speed, the actual shock is faster (See also Fig. \ref{figure2}(c)). Figure \ref{figure2}(c) reveals the reason for the different shock speed: the post-shock density of the plasma along $\rho = 90^\circ$ is $\approx 2.2$ while it is $\approx 3.3$ for an angle $\rho = 0$. The lower compression along $\rho = 90^\circ$ leads to a shock speed that is larger in the rest frame of the downstream ions.

After it reached a maximum at the end of the rarefaction wave the mean velocity of the dense beams of blast shell ions decreases until it reaches a minimum at $y=3.4$ mm in Fig. \ref{figure3}(a) and at $x=3.2$ mm in Fig. \ref{figure3}(b). Such a velocity decrease must be tied to an electric field. Figure \ref{figure2}(a) shows that the rapid initial decrease of the mean speed in Fig. \ref{figure3}(a) coincides with an increase of the density for $x=0$. The density increase at $x=2.7$ and $y=0$ in Fig. \ref{figure2}(a), which is a consequence of the slowdown of the blast shell ions, and that of the magnetic pressure in Fig. \ref{figure2}(b) result in an electric field which is tied to the different mobility of electrons and ions in gradients of the density and magnetic pressure. This electric field decreases the ion velocity at this value of $x$ in Fig. \ref{figure3}(b). 

\begin{figure}
\includegraphics[width=\columnwidth]{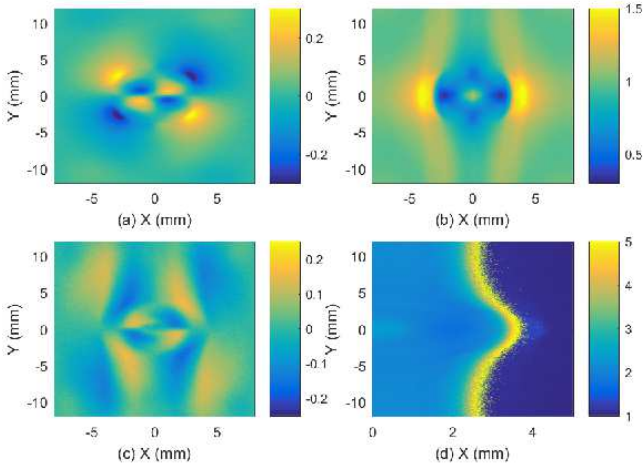}
\caption{The magnetic $B_x$ component is shown in (a), $B_y$ in panel (b) and $B_z$ in panel (c) in units of $B_0$. Panel (d) shows the mean kinetic energy per electron in units of the mean kinetic energy of electrons with the temperature $T_0$. The time is $t_{sim}=3.57$ ns (Multimedia view).}\label{figure4}
\end{figure}

Figure \ref{figure4} shows the amplitudes of the three magnetic field components and the mean kinetic energy of the electrons.
The magnetic $B_y$ component is the main contributor to the magnetic pressure $P_B$. The large amplitudes of $B_x$ are caused by the bending of the field lines by the plasma expansion along x. This can be seen from a comparison of Fig. \ref{figure4}(a,b), for example at the location $(x,y)\approx$ (3 mm, -2 mm). The positive value for $B_x$ implies that the magnetic field line bends at this location towards increasing $x$, which is consistent with Fig. \ref{figure4}(b). The same is true for the 4 magnetic field patches at small $r$ in Fig. \ref{figure4}(a). Their polarity is opposite to those at larger radii, because the magnetic field depletion due to the plasma expansion (See also Fig. \ref{figure2}(b)) lets the magnetic field lines move towards lower $r$. The magnetic $B_z$ component in Fig. \ref{figure4}(c) also shows structures that follow the deformation of the field lines in the x-y plane. The magnetic field deformation is accomplished by an electronic current. The electrons flow along the magnetic field lines to large $|y|$, which can be seen in Fig. \ref{figure4}(d). 

\begin{figure}
\includegraphics[width=\columnwidth]{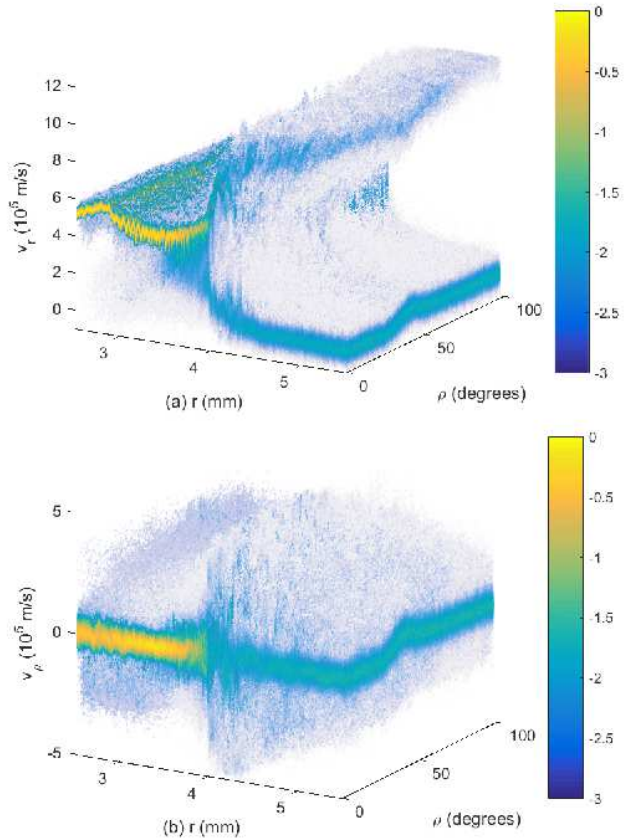}
\caption{The 10-logarithmic ion phase space density distribution. The time is $t_{sim}=3.57$ ns. Panel (a) shows the velocity in the radial direction $v_r$ while panel (b) shows the azimuthal velocity component $v_\rho$.}\label{figure5}
\end{figure}

Figure \ref{figure5} shows the phase space density distribution of the ions over the azimuthal interval $0 \le \rho \le 100^\circ$. It shows the distribution as a function of the radial velocity $v_r$ and the azimuthal velocity $v_\rho$.
The blast shell has driven a shock, which heats up the ambient ions, for all values of $\rho$. The blast shell ions form the dense core part in both distributions at $r<4$ mm. The thermal spread of the blast shell ions is larger in the azimuthal direction than in the radial direction; the reason being that the ions are cooled in the radial direction when the ambipolar electric field of the rarefaction wave accelerates them and when faster ions outrun the slower ones. The ambient ions, which have not yet encountered the shock, form the cool dense population at $r>$ 4 mm. The diffuse population for all $r$ is formed by ambient ions, which have crossed the shock ($r <$ 4 mm), and by shock-reflected ions ($r>$ 4 mm). 

A velocity modulation of the upstream ions is observed at $\rho \approx 45^\circ$ and $r\approx 5.4$ mm. Prior to the arrival of the shock the ions are accelerated to several times their thermal speed $v_{tn}={(k_BT_0/(12.5m_n))}^{1/2}\approx 3.3 \times 10^4$ m/s along the radial and azimuthal directions. This velocity increase along the radial direction persists up to a propagation direction $\rho = 100^\circ$ albeit with a lower magnitude. The shocks with propagation angles $45^\circ \le \rho \le 135^\circ$ thus have a foot, while those with $\rho \le 45^\circ$ have none.

\begin{figure}
\includegraphics[width=\columnwidth]{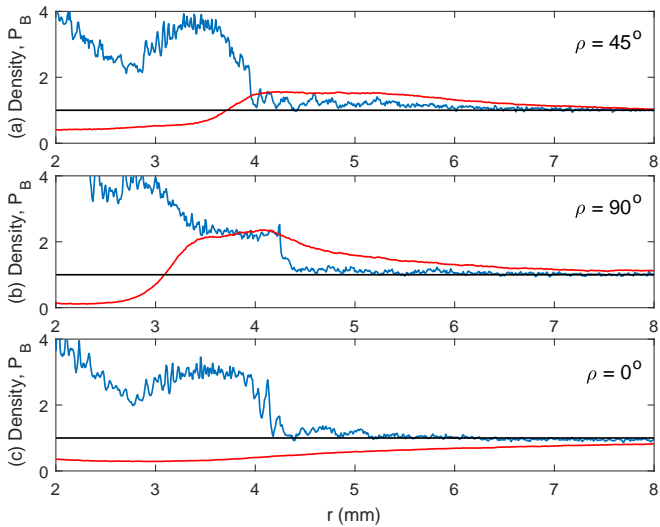}
\caption{The ion density (blue) and the magnetic pressure $P_B$ (red): panel (a) shows both along the direction $45^\circ$, panel (b) along the direction $90^\circ$ and (c) that along $\rho = 0$. Both distributions have been averaged over an angular interval with the width $0.5^\circ$. The time is $t_{sim}=3.57$ ns.}\label{figure6}
\end{figure}

Figure \ref{figure2}(c) showed that the high-density band reached the shock front at $\rho = 45^\circ$ and we want to assess its connection to the emergence of a shock foot. Figure \ref{figure6} compares slices of the ion density and of the magnetic pressure for the propagation angles $\rho = 45^\circ$ and $\rho = 90^\circ$ with those of the ion acoustic shock ($\rho = 0$).
The density distribution in Fig. \ref{figure6}(a) has a maximum between $r\approx 2.8$ mm and $r\approx 3.8$ mm. The peak density 3.6 downstream of the shock at $r\approx 3.5$ mm exceeds the post-shock peak density 2.2 at $r\approx 4$ mm in Fig. \ref{figure6}(b) and the maximum downstream density 3 of the shock in Fig. \ref{figure6}(c) at $r\approx 3.5$ mm. The magnetic pressure has been depleted at low radii in all considered cases. Figures \ref{figure6}(a,b) demonstrate that the shocks pile up magnetic field ahead of them. The magnetic pressure in Fig. \ref{figure6}(b) has started to increase even at the boundary at $r=8$ mm. 

A fast magnetosonic pulse emitted at $t=0$ at the boundary at $r=2$ mm would have reached the position $2 \, \mathrm{mm}+v_{fms}t_{sim} \approx 4.1$ mm in Fig. \ref{figure6}(b). The magnetic field is increased beyond this radius, which suggests together with the exponentially decreasing $P_B$ for $r>4.1$ mm that this pulse is a damped precursor. 

The magnetic pressure and the plasma density both increase in Fig. \ref{figure2}(b) as we cross the shock at $r \approx 4.3$ mm from the upstream into the downstream region. Based on this observation we have already concluded that it is a fast magnetosonic shock. The magnetic pressure decreases and the plasma density increases as we cross the shock at $r\approx 3.9$ mm in Fig. \ref{figure6}(a); it is a slow magnetosonic shock. The gradient of $P_B$ at the shock accelerates ions to lower radii, which enhances the plasma compression and yields the large post-shock density. 

The magnetic pressure gradient ahead of the slow- and fast magnetosonic shocks depicted in Fig. \ref{figure6}(a,b) points to increasing radii and the associated force accelerates ions in the same direction. The upstream ions in Fig. \ref{figure5} obtain a mean radial speed, which is larger than zero, that lets them move away from the shock. In contrast, the magnetic field is depleted ahead of the shock in Fig. \ref{figure6}(c) and its pressure gradient accelerates upstream ions towards the shock, which amplifies the mean velocity change at $\rho \approx 45^\circ$ in Fig. \ref{figure5}(a).

The depletion of $P_B$ in Fig. \ref{figure6}(c) extends far ahead of the shock and it can thus not be explained in terms of an Alfv\' en wave that is launched by the expanding blast shell at $r=2$ mm at $t=0$. The Alfv\'en speed is simply too low. Effects due to Alfv\'en waves and a modification of shocks by a shear Alfv\'en wave would also not emerge on the short time scales $t\ll \omega_{ci}^{-1}$ we consider here due to their low frequencies \cite{Gekelman11}. The magnetic field depletion can only be caused by the current of hot electrons, which is carried into the plasma at the electron's thermal speed $v_{te}\gg v_A$ (See Fig. \ref{figure4}(d)).

We have estimated the speed of the shocks shown in Fig. \ref{figure6} by measuring the distance the density jump associated with the forward shock crossed from $t=0.9t_{sim}$ until $t=t_{sim}$. The speed of the slow magnetosonic shock in Fig. \ref{figure6}(a) is about $4.5 \times 10^5$ m/s, that of the fast magnetosonic one in Fig. \ref{figure6}(b) is about $6.7 \times 10^{5}$ m/s or $\approx 1.15 v_{fms}$ and that of the unmagnetized shock in Fig. \ref{figure6}(c) is about $5.4 \times 10^5$ m/s or $1.3c_s$.

\subsection{Simulation 2: magnetic field aligned with z.}

\begin{figure*}
\includegraphics[width=\textwidth]{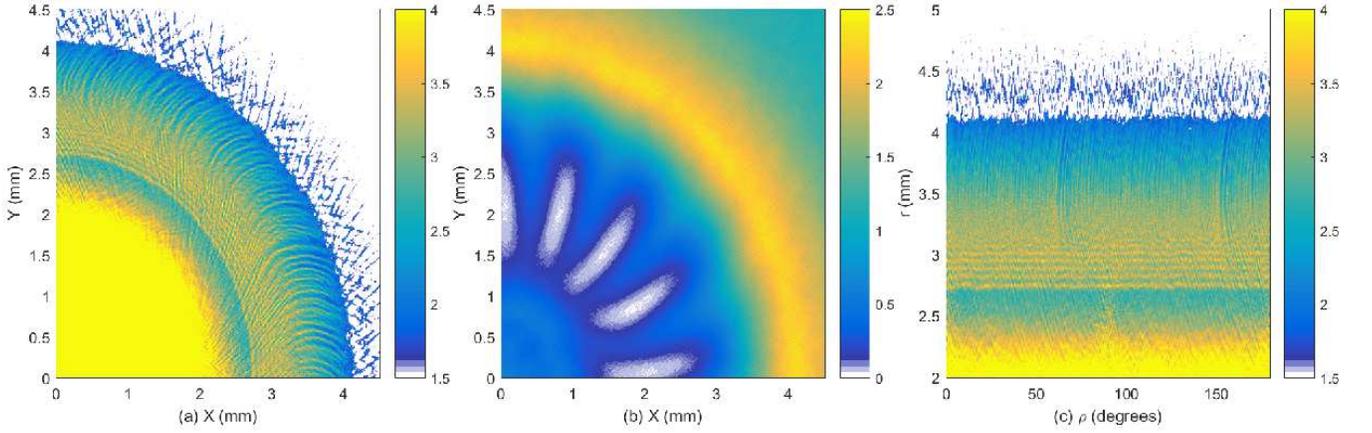}
\caption{The distribution of the ion density $n_i$ and of the normalized magnetic pressure $P_B=(B_x^2+B_y^2+B_z^2)/B_0^2$. The ion density is shown in cartesian coordinates in (a). Panel (b) shows the magnetic pressure in cartesian coordinates. The ion density distribution in polar coordinates is shown in panel (c). The linear color scale is clamped to the value 1.5 in (a, c) in order to emphasize the location of the shock. The time is $t_{sim}=3.57$ ns (Multimedia view).}\label{figure7}
\end{figure*}

Figure \ref{figure7}(a) shows the ion density in the quadrant $x>0$ and $y>0$. 
This distribution is radially symmetric for all radii with the exception of the density stripes found for $2.75\, \mathrm{mm} <r < 4\, \mathrm{mm}$. The magnetic pressure is radially symmetric only for $r\le 1.2$ mm and $r\ge 3.7$ mm as seen from Fig. \ref{figure7}(b). The magnetic pressure waves in the interval $1.3 \, \mathrm{mm} \le r \le 3.7\, \mathrm{mm}$ rotate in time in the counter-clockwise direction (Fig. \ref{figure7} multimedia view). Figure \ref{figure7}(c) shows that the plasma expansion is no longer a function of $\rho$. We observe the density stripes, which form at the end of the rarefaction wave at the same location $2.7 \mathrm{mm} \le r \le 3.2 \mathrm{mm}$, with the same amplitude and wavelength as their counterparts in Fig. \ref{figure2}(c) for $\rho=90^\circ$. Their amplitude and wavelength do not depend on $\rho$ because these waves are always lower-hybrid waves due to the orientation of the background magnetic field.

\begin{figure}
\includegraphics[width=\columnwidth]{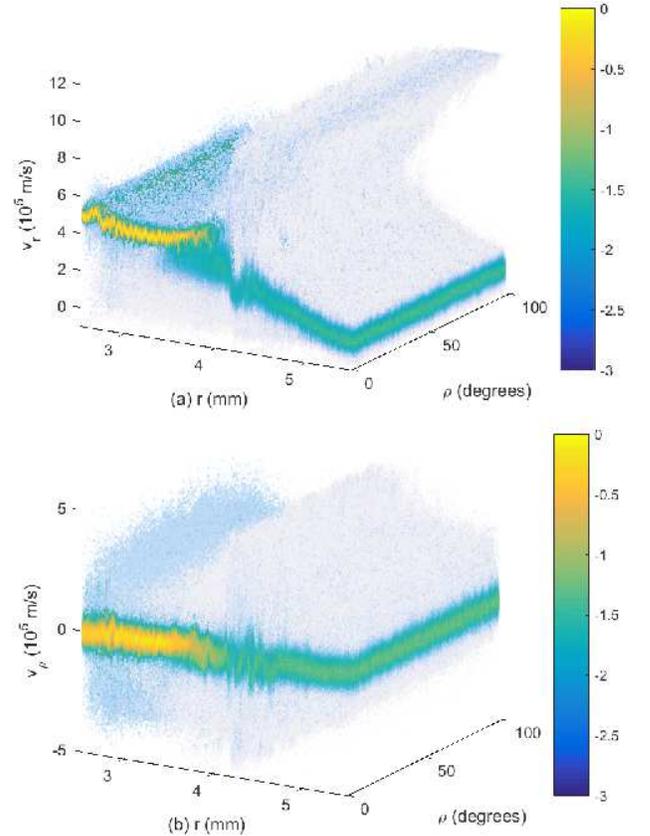}
\caption{The 10-logarithmic ion phase space density distribution. The time is $t_{sim}=3.57$ ns. Panel (a) shows the velocity in the radial direction $v_r$ while panel (b) shows the azimuthal velocity component $v_\rho$.}\label{figure8}
\end{figure}

The front of the shock in Fig. \ref{figure7}(a,b) is located at $r \approx 4$ mm and it compresses the plasma density and the magnetic pressure. It is a fast magnetosonic shock, which is underlined by the phase space density distribution of the ions in Fig. \ref{figure8}.
The distribution does not depend on $\rho$ and it resembles that in Fig. \ref{figure5} at $\rho = 90^\circ$. Figure \ref{figure8} furthermore reveals that the ion distribution has not been visibly affected by the azimuthal oscillations of the magnetic pressure in Fig. \ref{figure7}(b).

\begin{figure}
\includegraphics[width=\columnwidth]{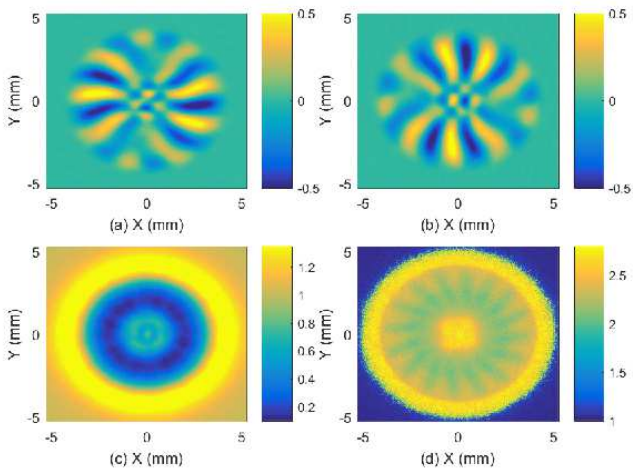}
\caption{The magnetic $B_x$ component is shown in (a), $B_y$ in panel (b) and $B_z$ in panel (c) in units of $B_0$. Panel (d) shows the mean kinetic energy per electron in units of those of electrons with the temperature $T_0$. The time is $t_{sim}=3.57$ ns (Multimedia view).}\label{figure9}
\end{figure}

Figure \ref{figure9} shows the distribution of the individual magnetic field components and the mean energy per electron at $t=t_{sim}$.
We observe modulations of the magnetic field components in the simulation plane and of the mean kinetic energy of the electrons. The wave vector of these oscillations is aligned with the azimuthal direction. 

The Weibel instability can lead to the growth of magnetic fields in a density gradient \cite{Quinn12}. However, it would affect the out-of-plane magnetic field, which is not what Fig. \ref{figure9} shows. The background magnetic field will also maintain a gyrotropic electron temperature in the simulation plane on the considered time scales $\omega_{ce}t_{sim}\approx 500$, which suppresses this instability. 

Instabilities can also be driven by a drift between electrons and ions. The guiding center approximation is valid for the electrons since they perform about 100 gyroperiods during $t_{sim}$. The radial change of $B_z$ in Fig. \ref{figure9}(c) leads to a grad-B drift. The drift velocity $\mathbf{v}_D$ of an electron with the charge $q=-e$ can be estimated by assuming that $B_x,B_y\ll B_z$ and that $\mathbf{B}=(0,0,B_z)$ changes slowly relative to the value of $B_z$ on spatial scales comparable to an electron thermal gyroradius. According to Fig. \ref{figure9}(c) this is the case for the electrons with $r_{ge}=1.25\times 10^{-4}$.  The drift speed is \cite{Treumann97}
\begin{equation}
\mathbf{v}_D \approx \frac{m_ev_{te}^2}{2qB_z} \frac{\mathbf{B} \times \nabla B_z}{B_z^2}.\label{driftspeed}
\end{equation}
Changes along $z$ are excluded by our simulation geometry. We define $c_D = (m_ev_{te}^2)/(2qB_z^3)$ and obtain the two drift components $v_{Dx}\approx  -c_DB_z\partial_yB_z$ and $v_{Dy} \approx  c_DB_z \partial_x B_z$. Electrons drift in the clockwise direction. 

\begin{figure}
\includegraphics[width=\columnwidth]{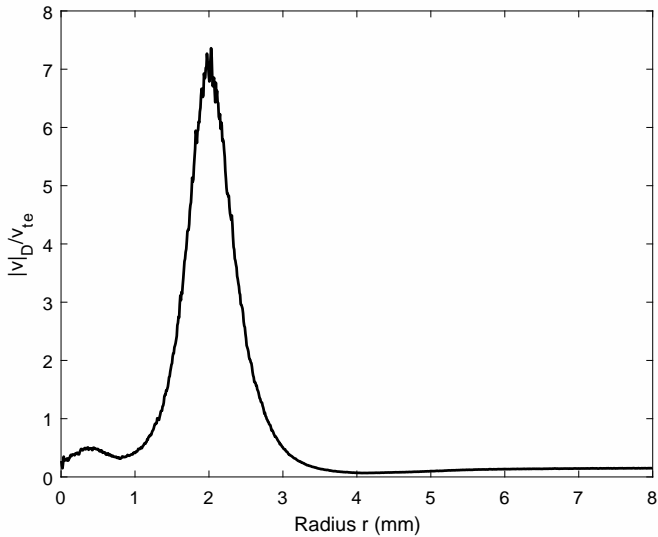}
\caption{The magnitude of the azimuthally averaged drift speed $|\mathbf{v}_D|/v_{te}$ at the time $t_{sim}=3.57$ ns.}
\label{figure10}
\end{figure}

We compute $\mathbf{v}_D$ with Eqn. \ref{driftspeed} from the magnetic field data in Fig. \ref{figure9}(c). The azimuthal average of its modulus is shown in Fig. \ref{figure10}.
The drift speed exceeds $v_{te}$ in the radial interval, in which we observe the magnetowaves in Figs. \ref{figure9}(a,b). Equation \ref{driftspeed} accurately estimates $\mathbf{v}_D$ only if $|\mathbf{v}_D|\ll v_{te}$ and beam instabilities will grow once $|\mathbf{v}_D|\sim v_{te}$. It is thus unlikely that electrons can reach a drift speed $7v_{te}$. This is confirmed by Fig. \ref{figure9}(d) that shows that the mean energy per electrons is only 3 times that of electrons with the temperature $T_0$.

Drift speeds below $v_{te}$ can drive the lower-hybrid drift instability \cite{Brackbill84,Winske88,Ripin93,Daughton04}, which results in ion density waves. The linear growth rate of these waves is below the lower-hybrid frequency $\omega_{lh}={({(\omega_{ci}\omega_{ce})}^{-1}+\omega_{pi}^{-2})}^{-1/2}$.
Figure \ref{figure9} (multimedia view) shows that the magnetowaves grow on time scales $\sim \omega_{ce}^{-1}$ while $\omega_{ce}/\omega_{lh} \approx 60$ in the ambient plasma. The slow growth of lower-hybrid waves and the absence of ion density modulations Fig. \ref{figure7}(a) on spatial scales that are similar those of the magnetic pressure in Fig. \ref{figure7}(b) rule out this instability. 

A faster-growing instability that involves electron-cyclotron waves \cite{Forslund72,Dieckmann00} sets in if the drift speed is comparable to the electron's thermal speed. Such waves hardly modulate the ion density. They are sustained by the interplay of the magnetic pressure with the electron thermal pressure when they saturated nonlinearly. The multimedia view of Fig. \ref{figure9}(d) evidences modulations of the mean thermal energy of the electrons, which suggests that such an instability is involved.  

\begin{figure}
\includegraphics[width=\columnwidth]{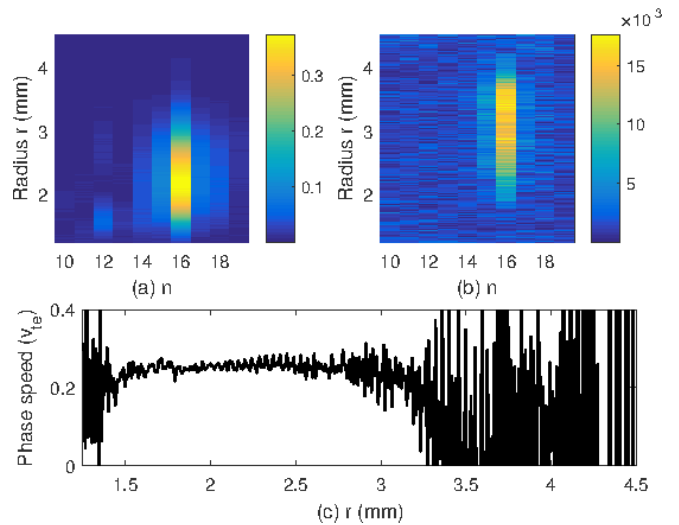}
\caption{Amplitude modulus and phase speed of the drift wave: panel (a) shows the amplitude modulus of the magnetic pressure and panel (b) that of the mean energy per electron. A value $n=1$ implies one oscillation per full circle. Both spectra are normalized for each value of $r$ to the power at $n=0$ at that $r$. Panel (c) shows the propagation speed of the magneto-structure as a function of the radius.}
\label{figure11}
\end{figure}

We can quantify a correlation between the magnetic pressure, which oscillates twice as fast as the magnetic amplitude in Fig. \ref{figure9}(a,b), and the mean electron energy by transforming both from a cartesian into a polar coordinate system followed by a Fourier transform over the azimuth angle. We define $n$ as the number of oscillations along the azimuthal direction; one oscillation per $360^\circ$ corresponds to $n=1$. The result is shown in Fig. \ref{figure11}(a, b).    
An almost monochromatic signal is observed at $n=16$. The signal in Fig. \ref{figure11}(a) peaks at a lower radius than that in Figure \ref{figure11}(b). The density decreases with increasing $r$ and the magnetic pressure must thus be balanced by a larger kinetic energy per electron. 

We have extracted the phase angle $\alpha_1(r)$ of the Fourier transformed signal at $n=16$ in Fig. \ref{figure11}(a) and that of its counterpart $\alpha_2(r)$ at $t=t_{sim}-54$ ps. A phase difference $\Delta_\alpha = \alpha_1(r)-\alpha_2(r) = 2\pi$ implies that the structure has completed one full rotation around the z-axis. The azimuthal phase speed of the structure $r\Delta_\alpha/54$ ps is shown in Fig. \ref{figure11}(c). A positive phase speed corresponds to a counter-clockwise rotation. The phase velocity equals $v_{te}/4$ over the radial interval $1.5 \, \mathrm{mm} \le r \le 3.3 \, \mathrm{mm}$, which is typical for electron Bernstein mode waves. The magnetowave rotates in the opposite direction as the electrons, which is typical for waves driven by the resonant electron-cyclotron drift instability. The opposite rotation direction is caused by the oppositely directed phase and group velocities of electron Bernstein mode waves. Energy is transported with the group velocity in the same direction as the electrons move and the phase velocity thus has the opposite sense of rotation. This effect has been observed experimentally \cite{Ripkin73}. 

\section{Summary}

We have examined the expansion of an initially radially symmetric blast shell into an ambient plasma and the ensuing formation of a magnetic cavity \cite{Bernhardt87} by means of two-dimensional particle-in-cell simulations. The blast shell was driven by a jump in the thermal pressure between a dense circular plasma cloud and a spatially uniform dilute ambient medium. A spatially uniform magnetic field permeated the plasma. 

One simulation considered the case of a magnetic field that was aligned with one of the axes of the simulation box. This simulation demonstrated that collisionless forward shocks form for all orientations of the shock normal with the ambient magnetic field. The forward shocks were located between the pristine and the shocked ambient medium. The forward shock that propagated perpendicularly to the magnetic field ($\rho = 90^\circ$) was mediated by the fast magnetosonic mode. It was trailed by a tangential discontinuity, which separated the blast shell plasma from the shocked ambient medium. The tangential discontinuity changed into a slow magnetosonic shock for propagation angles $45^\circ \le \rho < 90^\circ$ \cite{Eliasson14}. The slow magnetosonic shock replaced the fast magnetosonic shock as the forward shock for the propagation angle $\rho \approx 45^\circ$. The forward shocks were mediated by the ion acoustic wave for propagation angles $\rho < 45^\circ$. 

Our simulation time was only a few percent of an inverse ion gyro-frequency. In spite of this short time, the shocks developed features that are typical for magnetosonic shocks like the correct phase relation between the plasma density and the magnetic field amplitude and a shock speed that depended on the propagation direction. We attribute this fast formation time to the fact that the magnetosonic shocks are mediated by the magnetic pressure gradient force. The magnetic pressure gradient force acts on the electrons and ions. The higher mobility of the electrons results in a charge separation and, hence, in an electrostatic field. The characteristic time scale, during which the force develops, is between the ion and electron time scales and thus much shorter than the time it takes an ion to complete one gyration in the magnetic field. Indeed it has been shown in the one-dimensional simulation in Ref. \cite{Dieckmann17} that the shock, which propagates orthogonally to the magnetic field, becomes a fast magnetosonic one.

The second simulation considered a background magnetic field that pointed out of the simulation plane. This geometry implied that all forward shocks were mediated by the fast magnetosonic mode. No difference between these shocks and the perpendicular one in the first simulation was observed. The expanding blast shell depleted the magnetic field and piled it up ahead of it in the shocked ambient medium. The spatially non-uniform magnetic field resulted in the grad-B drift of electrons in the simulation plane at a speed that was comparable to the electron thermal speed. This drift speed was large enough to trigger the growth of an electron-cyclotron drift instability.

The distribution and evolution of the ion phase space density, which defined the collision-less shock, matched the one we would expect from an MHD model with the exception of the shock-reflected ion beam. The latter is absent in collisional plasmas. It can drive instabilities upstream of the shock or force it into a cyclic reformation. These kinetic effects were negligible in our simulation because the shock-reflected ion beam was dilute. We have not examined here the electrons; their high temperature implied that they were not in a thermal equilibrium with the ions. They provided the thermal pressure that led to the expansion of the blast shell and the current that deformed the magnetic field and resulted in the kinetic drift instability.

Future simulations have to examine the evolution of the shock for a wider range of values for $\beta$ and for the blast shell's expansion speed in order to determine cases for which the shock evolution starts to deviate from that predicted by an MHD model.

\textbf{Acknowledgements:} the simulations were performed on resources provided by the Swedish National Infrastructure for Computing (SNIC) at HPC2N (Ume\aa) and on resources provided by the Grand Equipement National de Calcul Intensif (GENCI) through grants x2016046960 and A0010506129. The EPOCH code has been developed with support from EPSRC (grant No: EP/P02212X/1). MED acknowledges support by a visiting fellowship of CRAL (ENS de Lyon). GS wishes to acknowledge support by EPSRC (grant No: EP/N027175/1). DF and RW thank the French National Program for High Energy (PNHE) for support. We thank the P\^{o}le Scientifique de Mod\'{e}lisation Num\'{e}rique - PSMN at ENS-Lyon where part of the analysis has been done.


\section*{References}
\begin{thebibliography}{10}

%


\bibitem{Forslund71} D. W. Forslund, and J. P. Freidberg, Phys. Rev. Lett. \textbf{27}, 1189 (1971).
\bibitem{Lembege01} B. Lembege, and F. Simonet, Phys. Plasmas \textbf{8}, 3967 (2001).
\bibitem{Sack87} Ch. Sack, and H. Schamel, Phys. Rep. \textbf{156}, 311 (1987).
\bibitem{Grismayer06} T. Grismayer, and P. Mora, Phys. Plasmas \textbf{13} 032103 (2006).
\bibitem{Grismayer08} T. Grismayer, P. Mora, J. C. Adam, and A. Heron, Phys. Rev. E \textbf{77}, 066407 (2008).
\bibitem{Thaury10} C. Thaury, P. Mora, A. Heron, J. C. Adam, and T. M. Antonsen, Phys. Rev. E \textbf{82}, 026408 (2010).

\bibitem{Dieckmann12} M. E. Dieckmann, G. Sarri, G. C. Murphy, A. Bret, L. Romagnani, I. Kourakis, M. Borghesi, A. Ynnerman, and L. O'C. Drury, New J. Phys. \textbf{14}, 023007 (2012).


\bibitem{Dieckmann16} M. E. Dieckmann, G. Sarri, D. Doria, A. Ynnerman, and M. Borghesi, Phys. Plasmas \textbf{23}, 062111 (2016).

\bibitem{Dieckmann17} M. E. Dieckmann, D. Folini, R. Walder, L. Romagnani, E. d'Humieres, A. Bret, T. Karlsson, and A. Ynnerman, Phys. Plasmas \textbf{24}, 094502 (2017). 

\bibitem{Lembege92} B. Lembege, and P. Savoini, Phys. Fluids B \textbf{4}, 3533 (1992).

\bibitem{Scholer92} M. Scholer, and D. Burgess, J. Geophys. Res. \textbf{97}, 8319 (1992).

\bibitem{Shimada00} N. Shimada, and M. Hoshino, Astrophys. J. \textbf{543}, L67 (2000).

\bibitem{Hoshino02} M. Hoshino, and N. Shimada, Astrophys. J. \textbf{572}, 880 (2002).

\bibitem{Scholer03} M. Scholer, I. Shinohara, and S. Matsukiyo, J. Geophys. Res. \textbf{108}, 1014 (2003).

\bibitem{Chapman05} S. C. Chapman, R. E. Lee, and R. O. Dendy, Space Sci. Rev. \textbf{121}, 5 (2005).

\bibitem{Lee05} R. E. Lee, S. C. Chapman, and R. O. Dendy, Phys. Plasmas \textbf{12}, 012901 (2005).

\bibitem{Burgess07} D. Burgess, and M. Scholer, Phys. Plasmas \textbf{14}, 012108 (2007).

\bibitem{Marcowith16} A. Marcowith, A. Bret, A. Bykov, M. E. Dieckman, L. O. Drury, B. Lembege, M. Lemoine, G. Morlino, G. Murphy, G. Pelletier, I. Plotnikov, B. Reville, M. Riquelme, L. Sironi, and A. S. Novo, Rep. Prog. Phys. \textbf{79}, 046901 2016.

\bibitem{Gueroult17} R. Gueroult, Y. Ohsawa, and N. J. Fisch, Phys. Rev. Lett. \textbf{118}, 125101 (2017).

\bibitem{Schaeffer17} D. B. Schaeffer, W. Fox, D. Habersberger, G. Fiksel, A. Bhattacharajee, D. H. Barnak, S. X. Hu, and K. Germaschewski, Phys. Rev. Lett. \textbf{119}, 025001 (2017).

\bibitem{Remington99} B. A. Remington, D. Arnett, R. P. Drake, and H. Takabe, Science \textbf{284}, 1488 (1999).

\bibitem{Zakharov03} Y. P. Zakharov, IEEE Trans. Plasma Sci. \textbf{31}, 1243 (2003). 

\bibitem{Bernhardt87} P. A. Bernhardt, R. A. Roussel-Dupre, M. B. Pongratz, G. Haerendel, A. Valenzuela, D. A. Gurnett, and R. R. Anderson, J. Geophys. Res. \textbf{92}, 5777 (1987).

\bibitem{Chapman89} S. C. Chapman, Planet. Space Sci. \textbf{37}, 1227 (1989).

\bibitem{Weibel59} E. S. Weibel, Phys. Rev. Lett. \textbf{2}, 83 (1959).

\bibitem{Quinn12} K. Quinn, L. Romagnani, B. Ramakrishna, G. Sarri, M. E. Dieckmann, P. A. Wilson, J. Fuchs, L. Lancia, A. Pipahl, T. Toncian, O. Willi, R. J. Clarke, M. Notley, A. Macchi, M. Borghesi, Phys. Rev. Lett. \textbf{108}, 135001 (2012).


\bibitem{Arber15} T. D. Arber, K. Bennett, C. S. Brady, A. Lawrence-Douglas, M. G. Ramsay, N. J. Sircombe, P. Gillies, R. G. Evans, H. Schmitz, A. R. Bell, and C. P. Ridgers, Plasma Phys. Controll. Fusion \textbf{57}, 113001 (2015).

\bibitem{Esirkepov01} T. Zh. Esirkepov, Comput. Phys. Commun. \textbf{135}, 144 (2001). 

\bibitem{Verscharen17} D. Verscharen, C. H. K. Chen, and R. T. Wicks, Astrophys. J. \textbf{840}, 106 (2017).

\bibitem{Dieckmann17b} M. E. Dieckmann, D. Doria, H. Ahmed, L. Romagnani, G. Sarri, D. Folini, R. Walder, A. Bret, and M. Borghesi, Phys. Plasmas \textbf{24}, 094501 (2017).

\bibitem{Gurevich84} A. V. Gurevich, and A. P. Meshcherkin, Eksp. Teor. Fiz. \textbf{87}, 1277 (1984).

\bibitem{Gekelman11} W. Gekelman, S. Vincena, B. Van Compernolle, G. J. Morales, J. E. Maggs, P. Pribyl, T. A. Carter, Phys. Plasmas \textbf{18}, 055501 (2011).

\bibitem{Treumann97} W. Baumjohann, and R. A. Treumann, \textit{Basic Space Plasma Physics} (Imperial College Press, 1997).

\bibitem{Ripin93} B. H. Ripin, J. D. Huba, E. A. McLean, C. K. Manka, T. Peyser, H. R. Burris, and J. Grun, Phys. Fluids B \textbf{5}, 3491 (1993).

\bibitem{Brackbill84} J. U. Brackbill, D. W. Forslund, K. B. Quest, and D. Winske, Phys. Fluids \textbf{27}, 2682 (1984).

\bibitem{Winske88} W. Winske, J. Geophys. Res. \textbf{93}, 2539 (1988).

\bibitem{Daughton04} W. Daughton, G. Lapenta, and P. Ricci, Phys. Rev. Lett. \textbf{93}, 105004 (2004).

\bibitem{Forslund72} D. Forslund, R. Morse, and C. Nielson, Phys. Fluids \textbf{15}, 1303 (1972).

\bibitem{Dieckmann00} M. E. Dieckmann, K. G. McClements, S. C. Chapman, R. O. Dendy, and L. O. C. Drury, Astron. Astrophys. \textbf{356}, 377 (2000).

\bibitem{Ripkin73} B. H. Ripkin, and R. L. Stenzel, Phys. Rev. Lett. \textbf{30}, 45 (1973).

\bibitem{Eliasson14} B. Eliasson, Phys. Plasmas \textbf{21}, 023111 (2014).
\end{thebibliography}
\end{document}